\documentclass[conference]{IEEEtran}
\IEEEoverridecommandlockouts
\newcommand{\thething}{reclaimID}
\usepackage{msc}
\usepackage{subcaption}
\usepackage{soul}
\usepackage[linesnumbered,ruled,lined]{algorithm2e}
\usepackage{flushend}
\usepackage[bookmarks=false,draft]{hyperref}
\def\msckeywordstyle{}

\mscset{/msc/msc keyword={}}
\usepackage{cite}
\usepackage{amsmath,amssymb,amsfonts}
\usepackage{algorithmic}
\usepackage{graphicx}
\usepackage{textcomp}
\def\BibTeX{{\rm B\kern-.05em{\sc i\kern-.025em b}\kern-.08em
    T\kern-.1667em\lower.7ex\hbox{E}\kern-.125emX}}
\begin{document}

\title{reclaimID: Secure, Self-Sovereign Identities using Name Systems and Attribute-Based Encryption\\
{}
}

\author{\IEEEauthorblockN{Blinded for review}}
\author{\IEEEauthorblockN{Martin Schanzenbach, Georg Bramm, Julian Sch\"utte}
\IEEEauthorblockA{\textit{Fraunhofer AISEC}\\
Garching near Munich, Germany \\
\{schanzen,bramm,schuette\}@aisec.fraunhofer.de}
}
%\and
%\IEEEauthorblockN{2\textsuperscript{nd} Georg Bramm}
%\IEEEauthorblockA{\textit{Fraunhofer AISEC}\\
%Garching near Munich, German \\
%bramm@aisec.fraunhofer.de}
%\and
%\IEEEauthorblockN{3\textsuperscript{rd} Dr. Julian Sch\"utte}
%\IEEEauthorblockA{\textit{Fraunhofer AISEC}\\
%Garching near Munich, Germany \\
%schuette@aisec.fraunhofer.de}
%}
\maketitle

  \begin{abstract}
In this paper we present \thething{}: An architecture that allows users to reclaim their digital identities by securely sharing identity attributes without the need for a centralised service provider. 
We propose a design where user attributes are stored in and shared over a name system under user-owned namespaces.
Attributes are encrypted using attribute-based encryption (ABE), allowing the user to selectively authorize and revoke access of requesting parties to subsets of his attributes.
We present an implementation based on the decentralised GNU Name System (GNS) in combination with ciphertext-policy ABE using type-1 pairings.
To show the practicality of our implementation, we carried out experimental evaluations of selected implementation aspects including attribute resolution performance. 
Finally, we show that our design can be used as a standard OpenID Connect Identity Provider allowing our implementation to be integrated into standard-compliant services.
    %{Please put abstract here.}
  \end{abstract}


\section{Introduction}
Today, users are often required to share personal data, like email addresses, to use services on the web.
%Such services often require user data when a particular action is executed (e.g. a mail is posted to the mailing list).
As part of normal service operation, such as notifications or billing, services require access to -- ideally fresh and correct -- user data.
Consider the use case of a user subscribing to a social networking service.
After successful registration and providing the service provider with an email address, the service uses it to send notifications such as status updates from friends.
At the time of notification delivery, the service needs access to the respective email addresses.
However, services cannot interact with users that are offline.

To mitigate this issue, services store user data in a database upon registration or retrieve it from a third party Identity Provider (IdP).
If the data is stored, it can become stale unless diligent users continuously update their data.
If the data is retrieved from an IdP, both user and service must be able to rely on the IdP to provide fresh, authentic attribute data.
Further, the IdP must be be available whenever needed and ideally does not abuse usage patterns, for example for user profiling.

Sharing user attributes in the Web today is often done via IdPs to reduce data redundancy and to give services access to current, up-to-date information even if the user is currently offline.
The most common approach is to use one of the two major IdPs: Google or Facebook.
Together they claim over 85\% of the identity provider market\footnote{\nolinkurl{http://www.gigya.com/blog/the-landscape-of-customer-identity-q2-2015/}, accessed 2017/02/20}.
The use of central service providers allows users to efficiently manage identity information and control access.
The IdP service is responsible for enforcing access control decisions made by the user regarding identities and attributes.
Consequently, the IdP has full access and control over the managed user data.
Abuse of this power is theoretically limited by local laws and regulations~\cite{Gola2005}.
But, they are often ignored or challenged~\cite{letter:EUdataprotfb} and centralised service providers are major targets for targeted advertisement businesses as well as hackers, including government actors~\cite{deibert2011access,gellman2013us,hui2013sina}.

%From a security perspective, the IdP is not only a single point of failure, but it also acts as an omniscient intermediary that learns all interactions between entities as they use the service.
From a security perspective, this setup of omniscient intermediaries is a significant threat to the users' privacy.
%as the IdP receives all attribute requests from all entities and thus learns about each interaction that involves any kind of authentication or authorization.
Users must completely trust the IdP with respect to protecting the integrity and confidentiality of their identity in their interest.
Various breaches of large IdPs such as the ones at Yahoo that revealed 3 billion user records to the public\footnote{\nolinkurl{https://en.wikipedia.org/w/index.php?title=Yahoo!_data_breaches&oldid=817379693}, accessed 2018/01/09} have shown that these expectations are hard to meet at times.
Finally, IdPs such as Facebook -- and for a long time also Google -- enforce a ``real-name policy''\footnote{\nolinkurl{http://www.businessinsider.de/facebook-changes-to-real-name-policy-2015-12}, accessed 2018/02/06}. Denying pseudononymous identity can be considered to be in direct violation to the human right to be forgotten.
%Further, applications rely on the availability of the IdP and an outage of globally used IdP such as Facebook would affect billions of users and hundreds of thousands of applications.

%The bottom line is that users have no guarantee that IdPs do not indeed analyse and market personal data from attributes and requests.
%One solution to this problem is to decentralise the Identity Provider services.
%Existing, recent approaches like NameID~\footnote{\nolinkurl{http://nameid.org}} or uPort~\footnote{\nolinkurl{https://www.uport.me}} decentralise data storage without providing adequate confidentiality protection mechanisms.

In this paper, we present the design and a reference implementation of \thething{}.
We address the issues elaborated above by not relying on a centralised IdP to serve attributes.
In \thething{}, users manage their attributes in a name system and can selectively grant other parties access.
The name system ensures that attributes can be accessed asynchronously whenever needed and provide integrity as well as authenticity guarantees.
Access to attributes is authorized and enforced through the use of attribute-based encryption (ABE).
We implemented this design using ciphertext-policy ABE and the GNU Name System~\cite{wachs2014censorship,wachs2014feasibility} to show that it can be practically realised.
Further, we show how \thething{} can be integrated into a standardised authorization and authentication protocol in the form of OpenID Connect.

\begin{figure*}[h!]
  \centering
    \begin{subfigure}[b]{0.4\textwidth}
      \includegraphics[width=\textwidth]{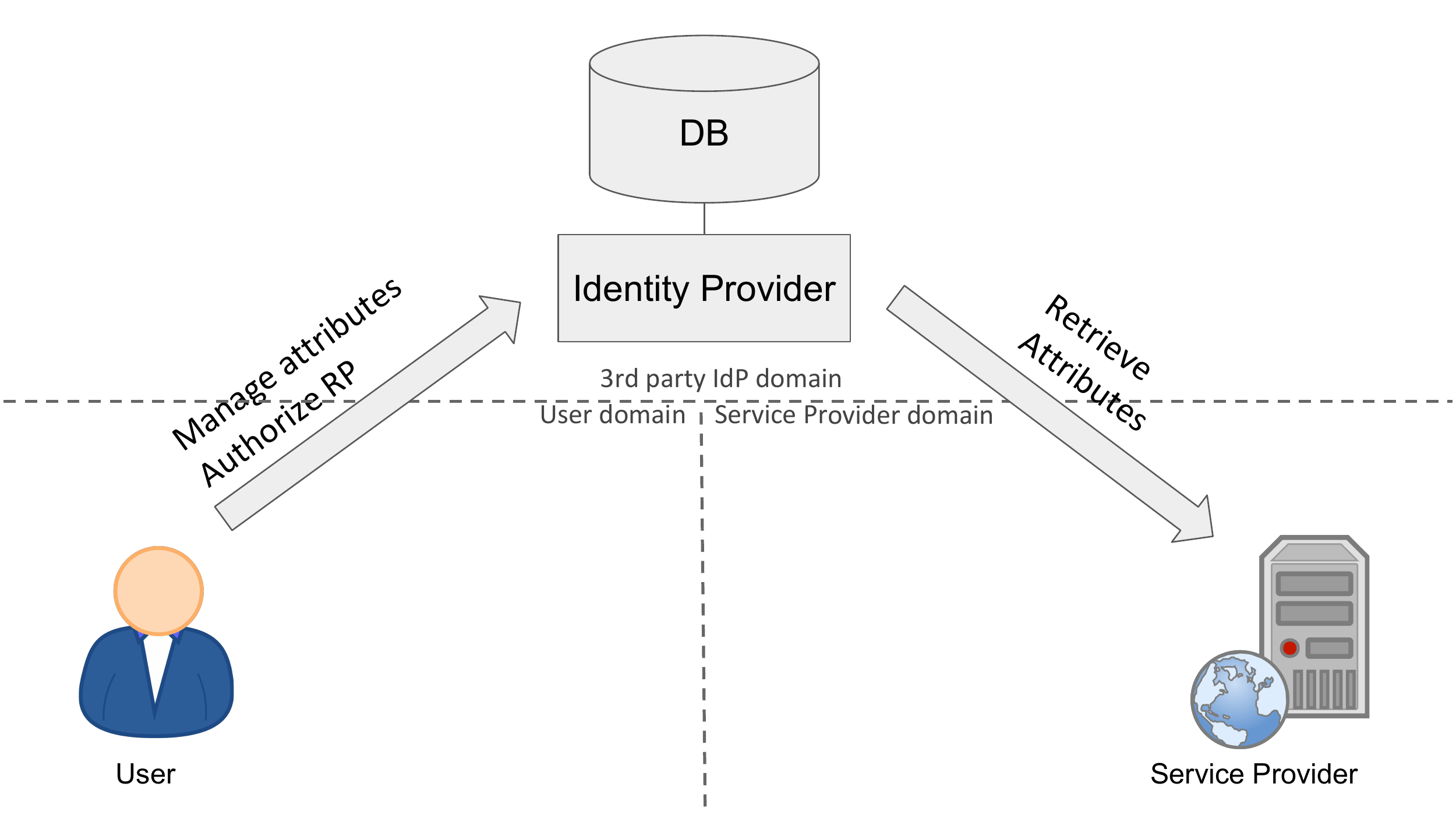}
      %\caption{Traditional, centralised Identity Provider setup. Common setup utilised by web application to access user attributes. }
      \caption{Centralised IdP, centralised storage}
      \label{fig:idp_traditional}
    \end{subfigure}
    ~ %add desired spacing between images, e. g. ~, \quad, \qquad, \hfill etc.
    \begin{subfigure}[b]{0.4\textwidth}
      \includegraphics[width=\textwidth]{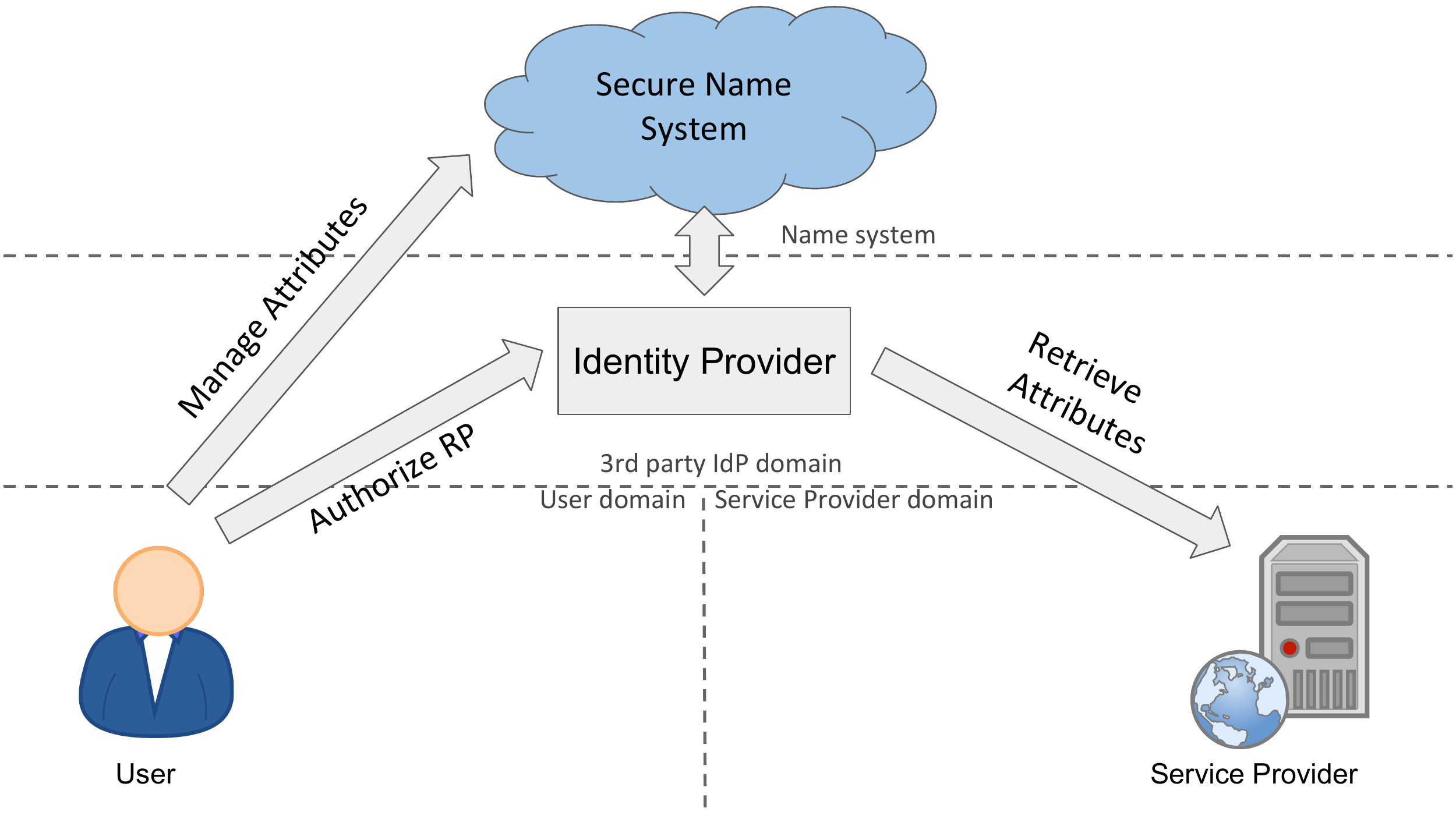}
      %\caption{Identity Provider with a decentralised user and attribute backend. While there is still a centralised service alloing management and access of identity attributes the actual information source is a decentralised name system.}
      \caption{Centralised IdP, decentralised storage}
      \label{fig:nameid}
    \end{subfigure}
    ~ %add desired spacing between images, e. g. ~, \quad, \qquad, \hfill etc.
    \begin{subfigure}[b]{0.4\textwidth}
      \includegraphics[width=\textwidth]{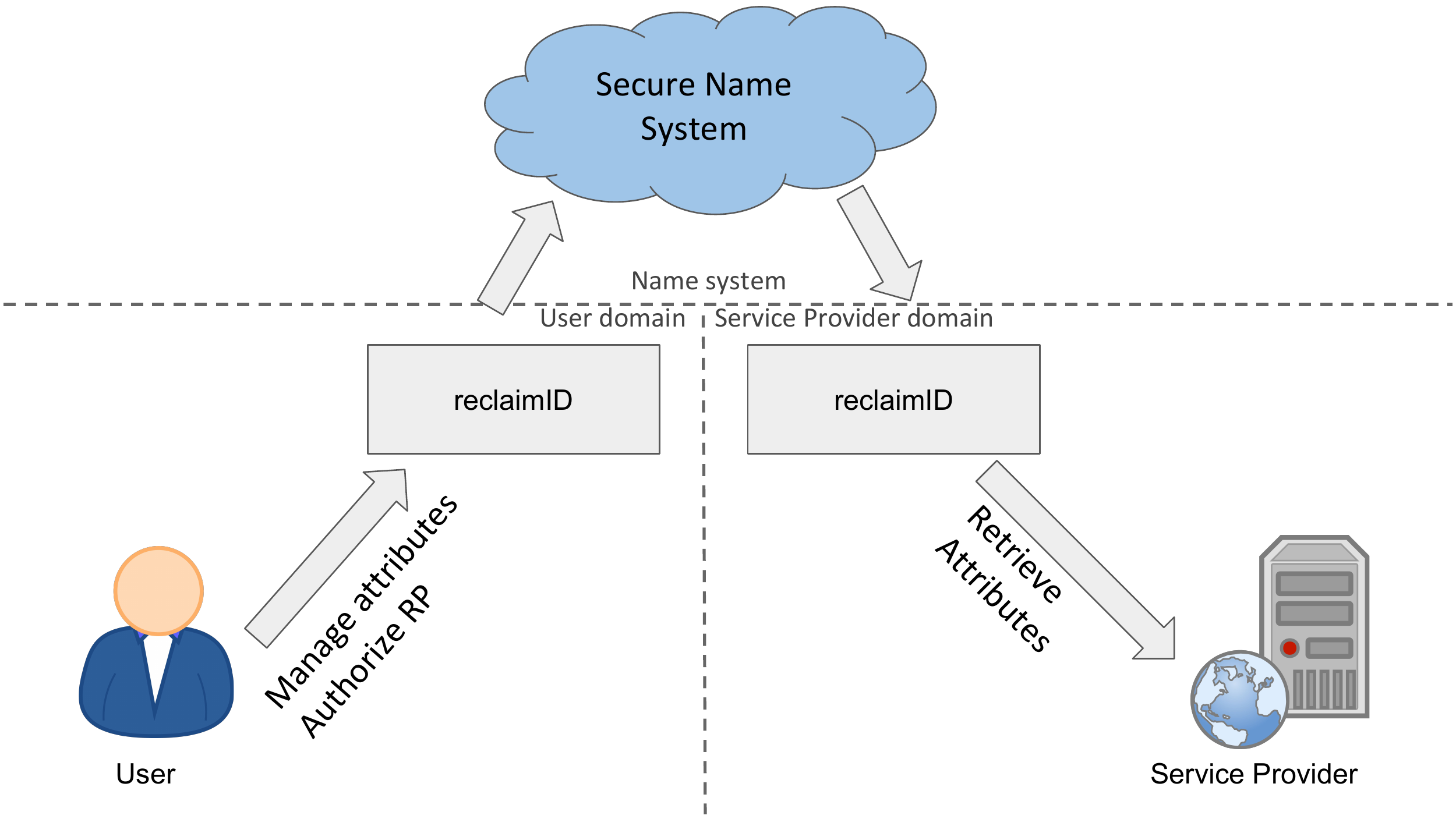}
      \caption{\thething{}, fully decentralised}
      \label{fig:gnuid}
    \end{subfigure}
    \caption{Identity Provider Architectures}
\end{figure*}

\section{Approach}

Traditionally, user attributes are centrally managed at an IdP and shared using protocols such as OpenID Connect 1.0 (OIDC) or SAML 2.0.
Requesting parties\footnote{In OpenID Connect referred to as ``relying party''} (RPs), as well as users, must trust the IdP with respect to availability, integrity, and confidentiality of attributes.
Figure~\ref{fig:idp_traditional} illustrates this setup.

An alternative to maintaining central user databases are decentralised approaches such as NameID~\cite{website:nameid}.
In NameID, there is a central IdP service that reads identity information from the \emph{Namecoin} name system.
Name systems consist of disjunct namespaces that contain mappings from human-readable names to values.
NameID equates namespaces with digital identities.
%In contrast to centralised managed user populations those identities and attributes are not asserted by a third party but are instead self-certified by the respective users claiming the identity attributes themselves.
%Such ``self-issued'' identities are by no means uncommon.
Figure~\ref{fig:nameid} illustrates the subtle difference between a identity management system with decentralised storage and traditional IdPs.
However, as even in this setup requests are still relayed over a single IdP, both architectures rely on a central IdP service that consequently acts as a single point of failure and is omniscient to all interactions between entities.
NameID additionally does not protect the confidentiality of information that is stored in Namecoin in any way.
%The design is equivalent to a world-readable database full of identities and attributes.
This is rendering the access control that a user may perform in the process of an authorization pointless.

The integration of identity management with recent name systems such as Namecoin or GNS has several advantages such as an out-of-the-box, close coupling of namespaces with cryptographic identities (i.e. public key pairs).
%In the case of the Domain Name System (DNS), this can only be achieved with the help of its respective security extensions: DNSSEC.
Further, name systems are not much different from identity management systems and most of their protocols can be used as a basis.
Specifically, we observe the following equivalences and differences between name systems and identity management (IdM) systems.

\begin{enumerate}
  \item Namespaces in name systems are equivalent to digital identities. Namespaces can be defined by or cryptographically associated with a public and private key pair.
  \item In a namespace, resource records are self-issued attributes of a user. These are equivalent to attribute names and values in an IdM system.
  \item Retrieving the attribute of an identity is equivalent to a name query in the respective namespace.
  \item Records in the name system can be queried by anyone. In an identity management system however, the confidentiality of identity attributes must be protected according to a user policy, i.e. when using a name system as the underlying layer of an IdM, such protection must be added on top.
\end{enumerate}

%As has been shown by Schanzenbach et al.~\cite{Schanzenbach2016}, it provides an exceptional set of security properties allowing an implementation of our design to circumvent censorship violations such as, for instance, service termination by the IdP due to real name policy violations.

With our approach, we set out to address the shortcomings discussed above.
By designing a decentralised identity management system that is user-managed and protects the users privacy we mitigate the requirement of a central IdP service that routes all requests.
We propose \thething{}, a decentralised IdP service that provides all typically required functions such as creation, querying, updates, and revocation of user identities in a decentralised way.
Similar to NameID, we use name systems as an underlying transport and storage medium.
However, \thething{} does not rely on any centralised component, as illustrated by Figure~\ref{fig:gnuid}.
The concept of \thething{} can be implemented on top of any name system, inheriting most of its security properties.
Additionally, \thething{} includes an additional authorization layer using ABE on top of the name system to ensure confidentiality and to enable policy-based access control of user attributes.

We design \thething{} to satisfy all requirements existing approaches also satisfy: From a user perspective it must be possible to manage one or more identities including respective attributes.
Further, the user must be able to selectively manage third party access to subsets of those attributes.
From a requesting party perspective it must be possible to request access to attributes of a user.
Those attributes must then be retrievable without a direct communication channel to the user.

\subsection{Security Goals and Threat Model}
Our goal is to ensure availability, authenticity, integrity, privacy and confidentiality of user attributes.
The user must be able to control access to his attributes by selectively authorising requesting parties.
As nation-wide manipulation of the domain name system, data leaks and the surveillance of global IdPs are a matter of fact today~\cite{deibert2011access,gellman2013us,hui2013sina}, our attacker model must consider an attacker with the ability to collect any data in transit between all participants, as well as manipulation of a limited but large number of nodes in the network.
We also assume that the attacker is able to coerce participants into submission of data they own or have access to.
However, we assume that the attacker is not able to break cryptographic primitives that are considered secure by the research community today. 
In the following we discuss the security properties \thething{} aims to satisfy in detail:

\textbf{Availability}
By not relying on a centralised service that serves user attributes and allows the user to manage authorizations, our system provides higher availability guarantees especially in the face of powerful attackers such as nation states that may utilise ``lawful'' interception techniques.
One major drawback of decentralised services with peer-to-peer connectivity is the possiblity of high user churn, which could potentially render identity attributes unavailable.
This would be problematic in use cases such as the social network provider, as the user attributes must be accessible by requesting parties even if the user is not online.
To solve this problem, our design is based on decentralising the service that allows users to manage attributes and selectively share them.
This is achieved by storing attributes in a name system where the user acts as the main authority over the data.
User attributes are stored in a distributed fashion and can be queried even if the user's peer is not online.
It should be noted that the concrete availability guarantees heavily depend on the underlying name system.
It is thus important to evaluate available name systems for an implementation of \thething{}.
We discuss name systems in Section~\ref{sec:implications} and our particular choice as part of the implementation in Section~\ref{sec:impl}.

\textbf{Authenticity and Integrity}
Users store attributes through the use of the IdP in a namespace of a name system.
We assume that by doing so, attributes are inherently self-signed using the private key associated with the identity of the user that owns the namespace.
Requesting parties can thus verify that the attribute was indeed stored by the user and has not been modified by any third party by simply verifying the signature.
%This usually happens implicitly when the name system is queried as no results are returned that did not pass a signature verification.
A more common use case is that a requesting party requires attributes that were issued by a trusted\footnote{trusted by the consumer of the attribute, e.g. the requesting party} third party.
In the case of Google or Facebook, the IdP is implicity the trusted issuer of all attributes.
Our proposed \thething{}, simlar to NameID, does not define the form factor of attributes and the user itself is implicity the issuer of attributes.
However, the use of attribute-based credentials (ABCs) is perfectly possible in both -- for example through the use of X.509 certificates as attributes.
%Apart from NameID, uPort~\cite{website:uPort} is another recent example in the emergence of identity management using distributed ledgers in combination with self-certified identity data.
%Even the OpenID Connect specification contains a brief section on self-issued ID tokens~\cite{website:oidc} and one can argue that IdPs like Google and Facebook also accept attributes that are, to some degree, self-certified by the users.

%\todo{I do not understand the following sentence}
%The association between the identity and the public key, as well as the transfer of this information to the requesting party is part of the authorization protocol.

\textbf{Privacy and Confidentiality}
We argue that given centralised IdP services, an attacker can coerce the service provider and then has knowledge over all connections between users and requesting parties.
The attacker can learn, for example, which services the user accesses over time.
Our design does not directly mitigate this issue as some name systems do not protect the confidentiality of queries and responses as well as metadata.
Name systems such as Namecoin and GNS can offer such protection due to built-in ``query privacy''.
We will discuss the security properties of selected name systems later.

To generally ensure confidentiality of attributes, we propose to encrypt them using ABE before they are stored in the name system.
%ABE comes in two main variants, Key-Policy (KP-ABE) and Ciphertext-Policy (CP-ABE).
Through the ABE-layer of \thething{}, requesting parties are issued user keys that only allow decryption of those attribute subsets that they are authorized to access.
We propose to boostrap an ABE system for every identity.
This means that every user acts as its own, individual key authority, giving him full and exclusive authority over all user keys and attributes.
This achieves both, confidentiality of user attributes in the otherwise public namespace of the name system, and fine-grained access control of requesting parties to these attributes.
Revocation of access rights to attributes is achieved by the creation of new keys, deletion of the existing attributes and publication of re-encrypted attributes over the name system.
A further advantage of enforcing access rights cryptographically through ABE rather than traditionally through a central trusted IdP is that individually encrypted attributes profit from the caching implemented in most name systems and significantly reduce network overhead for the retrieval of attributes (even to zero, if the cache is local).
We note that as our attacker is able to coerce participants into submission of data this approach does not protect against the case where an authorized third party is attacked.
However, unlike an IdP, an authorized third party likely does not have access to \emph{all} the data of a single user and cannot observe access patterns of other services.

Another aspect to privacy is that a user might not want to disclose the attribute value to a requesting party.
Instead, only certain properties should be disclosed.
A common use case is age verification, where a user only wants to disclose that she is over a certain age.
Advancements in the area of privacy-preseving attribute-based credentials (PP-ABCs)~\cite{cryptoeprint:2014:468,camenisch2002design} allow this through the use of interactive zero-knowledge proofs.
In our design the interactivity requirement is not acceptable, but recent work in the form of non-interactive zero-knowledge proofs~\cite{ben2013snarks} can be used in \thething{}, if needed.

\section{Design}
In this section, we present the main design of \thething{}'s protocols.
We show how we combine a name system and ABE scheme to realise a privacy-preserving, decentralised IdP.
In particular, we answer the question of how users can grant and revoke authorizations for requesting parties to access their attributes. 
Further, we discuss the protocol for authorized parties to retrieve and decrypt user attributes from \thething{}.
Finally, we elaborate on the impact that the choice of name system as well as ABE scheme has on an implementation of \thething{}.

\subsection{Preliminaries}
Attribute-based encryption schemes come in the two main flavours of ciphertext-policy ABE (CP-ABE) and key-policy ABE (KP-ABE).
As our access policies refer only to single labels in the name system, both variants are likewise suited.
For a first implementation we chose CP-ABE. 
%TODO UNCOMMENT TO INCLUDE KP-ABE
%KP-ABE counterparts for relevant functions can be found in Appendix~\ref{app:kpabe}.
A thorough discussion on the various implications of the two types of ABE schemes and the various existing schemes that exist for them can be found in Section~\ref{sec:implications}.
For the discussion of the system's design, both variants can be considered equally possible.
In the following, we present three major components: A name system, a CP-ABE scheme and an IdP.
We explain how in \thething{} the first two components are used to realise the third.

First, we define the high-level functions and procedures and objects for all components.
Let an ABE scheme consist of the following functions:
\begin{equation}
  \begin{split}
    \mathbf{Setup_{\mathsf{ABE}}}() &\rightarrow (msk_{\mathsf{ABE}}, pk_{\mathsf{ABE}}) \\
    \mathbf{Keygen_{\mathsf{ABE}}}(msk_{\mathsf{ABE}}, A) &\rightarrow sk_{\mathsf{ABE}}\\
    \mathbf{Enc}_{\mathsf{ABE}}(pk_{\mathsf{ABE}}, pt, policy) &\rightarrow ct\\
    \mathbf{Dec}_{\mathsf{ABE}}(sk_{\mathsf{ABE}}, ct) &\rightarrow pt
  \end{split}
\end{equation}
Where $msk_{\mathsf{ABE}}$ is the master secret key, $pk_{\mathsf{ABE}}$ the public parameters key and $sk_{\mathsf{ABE}}$ a derived user key in the $ABE$ scheme. 
$A$ is a set of tags, or attribute names that can be associated with a key $sk_{\mathsf{ABE}}$ using the function $\mathbf{Keygen_{\mathsf{ABE}}}()$.
Here, $policy$ describes the policy that is attached to a ciphertext $ct$.
Finally, $pt$ denotes the plaintext message.
For encryption and decryption we define the functions $\mathbf{Enc}_{\mathsf{ABE}}()$ and $\mathbf{Dec}_{\mathsf{ABE}}()$, respectively.

For the name system and IdP, we define $pk_{\mathsf{user}}, sk_{\mathsf{user}}$ as the public and private key pair associated with an identity $ID_{\mathsf{user}}$.
We define the functions $\mathbf{Enc}()$ and $\mathbf{Dec}()$ as the asymmetric encryption and decryption functions for use with the identity key pair.

Name systems consist of \emph{namespaces} that are owned by users or legal entities.
Namespaces are managed by their respective owners and contain name-value mappings. 
In our use case, the owner of a namespace is an ``issuer'' of attributes in its respective namespace. 
As such, name systems inherently provide a storage, resolution and delegation
mechanism for self-issued attributes.
Such name-value mappings are realised in name systems using \emph{resource
records}.
In name systems this mapping must be cryptographically bound to the namespace
owner, usually through digital signatures using public-key cryptography.
Let a name system $N$ consist of the following procedures
\begin{equation}
  \begin{split}
    \mathbf{Resolve} (ID_{\mathsf{user}}, name) &\rightarrow R\\
    \mathbf{Publish}(ID_{\mathsf{user}}, name, R)\\
    \mathbf{Depublish}(ID_{\mathsf{user}}, name)
  \end{split}
\end{equation}
$R$ is a record in a namespace and $name$ is a name in a namespace owned by $ID_{\mathsf{user}}$.
While it may seem odd that a resolution takes a name \emph{and} a namespace as argument it is not.
Internally resolvers in a name system always query for names inside namespaces.
It is common, however, that the resolver has specialised procedures that hide this from the user, e.g. by performing iterative lookups and segmenting a long name into multiple labels.
The prime example here is DNS, which allows a user to resolve a ``fully qualified domain name'' (FQDN) by iteratively trying to find the authoritative namespace for a specific label.
We assume that in the name system $sk_{\mathsf{user}}$ is used to create a record signature over the data in a record $R$ in a namespace owned by $ID_{\mathsf{user}}$.
When $R$ is published using $\mathbf{Publish}()$, the signature is stored alongside the record.
Consequently, we also assume that the record signature is verified when a record is retrieved using $\mathbf{Resolve}()$.
The respective public key can be used to uniquely identify the namespace and verify record signatures.
A published record is no longer resolvable after calling $\mathbf{Depublish}()$ and the cached records expire.

Finally, let an $IdP$ consist of the procedures:
\begin{equation}
  \begin{split}
    \mathbf{Store} (ID_{\mathsf{user}}, attribute)\\
    \mathbf{Delete}(ID_{\mathsf{user}}, attribute)\\
    \mathbf{Authorize}(ID_{\mathsf{user}}, ID_{\mathsf{rp}}, attributes) &\rightarrow ticket\\
    \mathbf{Revoke}(ticket)\\
    \mathbf{Retrieve}(ID_{\mathsf{rp}}, ticket) &\rightarrow attributes\\
  \end{split}
\end{equation}
The procedures $\mathbf{Store}()$ and $\mathbf{Delete}()$ allow the user $ID_{\mathsf{user}}$ to manage attributes.
$\mathbf{Authorize}()$ is the procedures used to authorize a requesting party $ID_{\mathsf{rp}}$ to access a set of attributes.
This access can be revoked using $\mathbf{Revoke}()$.
The requesting party can use the $\mathbf{Retrieve}()$ procedures to access attributes it was granted access to.

We define an identity $attribute$ as follows:
\begin{equation}
  \begin{split}
    attribute = (name, value, version)
  \end{split}
\end{equation}
The $name$ is an attribute identifier, such as ``email''. 
An attribute also has a $value$ associated with it.
The $value$ may contain arbitrary data associated with $name$ such as ``john@doe.com''.
It may also contain more complex data structures such as credentials issued by third parties.
The details of attribute values, however, are out of scope in our design.
The attribute $version$ is relevant for revocation in the later sections of this chapter.
%When access to an attribute is revoked, the respective record must be re-encrypted using a different tag as to not allow holders of ABE keys containing the old tag to access the value.
%For this, the record is re-encrypted after incrementing the attribute version resulting in a different tag in the policy.

The $attributes$ specified in $\mathbf{Authorize}()$ and $\mathbf{Retrieve}()$ are a set of attributes. %JS: collection=set? Conjunction?
A $ticket$ is a handle of an authorization that is passed to the authorized requesting party so it can access the shared attributes.
We define a ticket as follows:
\begin{equation}
  \begin{split}
    ticket = (ID_{\mathsf{user}}, ID_{\mathsf{rp}}, names, rnd)
  \end{split}
\end{equation}
The ticket identities $ID_{\mathsf{user}}$ and $ID_{\mathsf{rp}}$ identify the user that issued the ticket and the requesting party, respectively.
$names$ is the list of attributes that the requesting party is authorized to access and 
$rnd$ is a random label under which the user key $sk_{\mathsf{ABE}}$ for the requesting party is stored encrypted in the namespace of the identity.
This ticket must be transferred in an initial out-of-band authorization process and is used by the requesting party to retrieve attribute data.

In the following, we always assume that given an identity, its public key $pk_{\mathsf{user}}$ and the associated ABE key material can also be retrieved.
If a procedure is called by an identity, we also assume that we have access to the respective private keys $sk_{\mathsf{user}}$, $sk_{\mathsf{rp}}$ and $sk_{\mathsf{ABE}}$.
Before storing the first attribute, a user must bootstrap an ABE system.
In this process, the user creates an ABE public parameters key $pk_{\mathsf{ABE}}$ and master secret key $msk_{\mathsf{ABE}}$ for one of her namespaces by executing $\mathbf{Setup_{\mathsf{ABE}}}()$.

\subsection{Storage}
In \thething{}, the encrypted value of an attribute is stored inside a resource record $R$ in the name system.
By publishing the resource record under the attribute name the user effectively issued an attribute to her identity.
In Algorithm~\ref{algo_publish} we define the IdP $\mathbf{Store}()$ procedure.

\IncMargin{1em}
\begin{algorithm}
  \SetKwData{Left}{left}\SetKwData{This}{this}\SetKwData{Up}{up}
  \SetKwFunction{Union}{Union}\SetKwFunction{FindCompress}{FindCompress}
  \SetKwInOut{Input}{input}\SetKwInOut{Output}{output}
  \Input{User attribute $\mathbf{a}$\\User identity $ID_{\mathsf{user}}$}
  %\Output{-}
  \BlankLine
  $policy$ $\leftarrow$ $\mathbf{Concat}(a.name, a.version)$\;
  $ct$ $\leftarrow$ $\mathbf{Enc}_{\mathsf{ABE}} (pk_{\mathsf{ABE}}, a.value, policy)$\;
  $\mathbf{Publish}(ID_{\mathsf{user}}, a.name, ct)$\;
  \caption{Store}\label{algo_publish}
\end{algorithm}\DecMargin{1em}

First, we use the concatenation procedure $\mathbf{Concat}()$ to build the ABE $policy$ from the attribute name and version.
The resulting policy can be interpreted as ``To decrypt the ciphertext, a key associated with a tag representing the attribute in the respective version is required''.
To create the record data that is stored in the name system, we encrypt the attribute value using the ABE encryption function $\mathbf{Enc}_{\mathsf{ABE}}()$.
The encrypted attribute value is published as a record under the attribute name using the name system function $\mathbf{Publish}()$.
We note here that internally name systems distinguish between different types of records.
We therefore define the record type of records representing identity attributes to be ``ID\_ATTR''.
The record type does not serve any specific function except from allowing us to distinguish our records from, e.g. IP addresses.
In our design, all attribute resource records must have this type set.

We also note here that records in name systems expire.
An implementation must make a choice for an appropriate expiration time that allows to efficiently make use of the respective caching mechanism in the name system, if any.
%An example wire format of an identity record in the name system can be found in Appendix~\ref{app:wireformats}.

\subsection{Authorization}
\label{sec:authorization}
To authorize a requesting party to access a set of attributes, the user must create an authorization-specific user secret key $sk_{\mathsf{ABE}}$ using the ABE function $\mathbf{Keygen_{\mathsf{ABE}}}()$.
For $sk_{\mathsf{ABE}}$ to be used to decrypt the respective attribute records of the shared attributes it must be associated with a specific set of \emph{tags}.

There are two ways an authorized party can learn $sk_{\mathsf{ABE}}$: Resolving it through the name system or via an out-of-band exchange, for example using a web-based authorization protocol.
The latter is only possible in ``synchronous'' use-cases, i.e. when user and authorized party are both online.
In use-cases where user or authorized party are offline, $sk_{\mathsf{ABE}}$ must be exchanged via the name system.
We define the procedure for authorization in Algorithm~\ref{algo_authorize}.

\IncMargin{1em}
\begin{algorithm}
  \SetKwData{Left}{left}\SetKwData{This}{this}\SetKwData{Up}{up}
  \SetKwFunction{Union}{Union}\SetKwFunction{FindCompress}{FindCompress}
  \SetKwInOut{Input}{input}\SetKwInOut{Output}{output}
  \Input{User identity $ID_{\mathsf{user}}$\\
         requesting party $ID_{\mathsf{rp}}$\\
         Set of attributes $A$\\
         Master secret key $msk_{\mathsf{ABE}}$\\}
  \Output{A ticket $t$}
  \BlankLine
  $tags$ $\leftarrow$ $\{\textbf{Concat}(a.name,a.version) \mid a \in A\}$\;
  $sk_{\mathsf{ABE}}$ $\leftarrow$ $\mathbf{Keygen}(msk_{\mathsf{ABE}}, tags)$\;
  $ct$ $\leftarrow$ $\mathbf{Enc}(pk_{\mathsf{rp}}, sk_{\mathsf{ABE}})$\;
  $rnd$ $\leftarrow_R$ $\mathbb{R}$\;
  $\mathbf{Publish}(ID_{\mathsf{user}}, rnd, ct)$\;
  $names$ $\leftarrow$ $\{a.name \mid a \in A\}$\;
  $t$ $\leftarrow$ $(ID_{\mathsf{user}}, ID_{\mathsf{rp}}, names, rnd)$\;
  \KwRet{$t$}\;
  \caption{Authorize}\label{algo_authorize}
\end{algorithm}\DecMargin{1em}
First, we generate a set of $tags$ that correspond to the respective encrypted records the requesting party shall be authorized to access.
After the $sk_{\mathsf{ABE}}$ is generated using the users' $msk_{\mathsf{ABE}}$, it is encrypted using the public key $pk_{\mathsf{rp}}$ of the requesting party.
Then, a random label $rnd$ is generated under which the encrypted $sk_{\mathsf{ABE}}$ is published in the user namespace.
The random label $rnd$, the user identity $ID_{\mathsf{user}}$, the requesting party identity $ID_{\mathsf{rp}}$ and the attributes that the requesting party is authorized to access are assembled into a ticket $t$.
Updates to $sk_{\mathsf{ABE}}$, made necessary for example due to revocations, are published by the user and retrieved by the requesting party using the same random label $rnd$.
Similarly to attribute records, we define key records to have a unique type of ``ABE\_KEY''.
%\todo{wire format?}

\subsection{Deletion}
Removing attributes is not as simple as removing the respective records from the namespace.
First, the attribute record may still be resolvable in the name system until the records expire and it is purged from the cache.
Requesting parties that are authorized to access this attribute then must be prohibited from accessing any future incarnations of this attribute. 
This is important as to not risk any unwanted side-effects where unauthorized parties may still be able to decrypt the attribute.
For this, the attribute tag version must be incremented before a new attribute with the same name is issued.
A \thething{} implemenation must keep track of this state by either keeping the attribute with an empty placeholder value or by having a local database that contains the versioning information.
This implementation detail is out of scope of the \thething{} design and we only define the procedure for deletion itself in Algorithm~\ref{algo_delete}.

\IncMargin{1em}
\begin{algorithm}
  \SetKwData{Left}{left}\SetKwData{This}{this}\SetKwData{Up}{up}
  \SetKwFunction{Union}{Union}\SetKwFunction{FindCompress}{FindCompress}
  \SetKwInOut{Input}{input}\SetKwInOut{Output}{output}
  \Input{User attribute $\mathbf{a}$\\User identity $ID_{\mathsf{user}}$}
  %\Output{-}
  \BlankLine
  $\mathbf{Depublish}(ID_{\mathsf{user}}, a.name)$\;
  $a.version++$\;
  \For{each ticket $t$ issued by $ID_{\mathsf{user}}$}{
    $A_{t}$ $\leftarrow$ $\{x \mid x \in A \setminus a \wedge x.name \in t.names\}$\;
       $\mathbf{Authorize}(ID_{\mathsf{user}}, t.ID_{\mathsf{rp}}, A_{t}, msk_{\mathsf{ABE}})$\;
     }
  \caption{Delete}\label{algo_delete}
\end{algorithm}\DecMargin{1em}
The $\mathbf{Delete}()$ procedure starts off by de-publishing the respective attribute record from the namespace and then incrementing the attribute version.
After, all authorized parties (i.e. all issued tickets) that have access to this attribute are re-authorised to access all attributes they had access to before \emph{except} the deleted attribute.

\subsection{Update}
When the user modifies the attribute value the respective record in the name system must be updated accordingly.
Naively, it is possible to simply combine a $\mathbf{Delete}()$ and a $\mathbf{Store}()$ call.
But since we defined the $\mathbf{Delete}()$ procedure to increment the attribute version such an approach would require the user to reissue all existing ABE keys to the relevant requesting parties.
Consequently, updating the attribute is simply a call to $\mathbf{Store}()$ after updating the attribute value.
This update will only take effect \emph{after} the identity record expires and only then will the updated value be resolvable by authorized parties.
As the tag used to encrypt the attribute does not change, previously authorized requesting parties will be able to decrypt the updated record data with their existing keys.

\subsection{Retrieval}
\label{sec:retrieval}
To retrieve an attribute $a$ of identity $ID_{\mathsf{user}}$ an authorized requesting party $ID_{\mathsf{rp}}$ must perform a lookup in the name system.
The name to lookup is the attribute name, e.g. ``email''.
If the attribute exists, the response from the name system will contain the encrypted attribute value record $R$.
As elaborated above, $sk_{\mathsf{ABE}}$ contains a set of tags that allows it to be used in the decryption of all attribute records that $ID_{\mathsf{rp}}$ is authorized to access.
To retrieve $sk_{\mathsf{ABE}}$, $ID_{\mathsf{rp}}$ must first resolve the key record under the name $rnd$ in the identity namespace of $ID_{\mathsf{user}}$.
To do so, $ID_{\mathsf{rp}}$ must have received the label $rnd$ out-of-band in a ticket as discussed in the previous section.
Given $sk_{\mathsf{ABE}}$, the requesting party can decrypt the attribute value using the CP-ABE decryption function $\mathbf{Dec}_{\mathsf{ABE}}()$.
We formally define the procedure for retrieval in Algorithm~\ref{algo_retrieve}.

\IncMargin{1em}
\begin{algorithm}
  \SetKwData{Left}{left}\SetKwData{This}{this}\SetKwData{Up}{up}
  \SetKwFunction{Union}{Union}\SetKwFunction{FindCompress}{FindCompress}
  \SetKwInOut{Input}{input}\SetKwInOut{Output}{output}
  \Input{requesting party $ID_{\mathsf{rp}}$\\
         Ticket $t$}
            %\Output{List of attributes}
  \BlankLine
  $ct$ $\leftarrow$ $\mathbf{Resolve} (t.ID_{\mathsf{user}}, t.rnd)$\;
  $sk_{\mathsf{ABE}}$ $\leftarrow$ $\mathbf{Dec}(sk_{\mathsf{rp}}, ct)$\;
  \For{all attribute names $n \in t.names$}{
    $R$ $\leftarrow$ $\mathbf{Resolve}(t.ID_{\mathsf{user}}, n)$\;
       $attribute$ $\leftarrow$ $\mathbf{Dec}_{\mathsf{ABE}}(sk_{\mathsf{ABE}}, R)$\;
     }
  \caption{Retrieve}\label{algo_retrieve}
\end{algorithm}\DecMargin{1em}
Note that most name systems allow queries for attribute records to be executed in parallel, which allows the for-loop in the $\mathbf{Retrieve}()$ procedure to be parallelised.

\subsection{Revocation}
%\todo{Also discuss identity revocation}
We define revocation -- as opposed to deletion -- as the process to revoke access of a specific requesting party to user attributes in \thething{}.
Revocation schemes for ABE are often quite complex and inefficient. %TODO citation needed
In our case we also have to take into account user key distribution and name system properties.

In fact, the performance impact caused by cryptographic operations is not as critical in our design for two reasons: 
First, regeneration of keys and re-encryption can be done locally in the background after it is initiated by the user. 
Second, from a requesting party point of view, even if access to a particular attribute is revoked there was a time in past where access was granted.
So, revoking access on currently accessible data is not important in our design.

Revocation of access in \thething{} is used to prevent the decryption of an attribute record using a specific user key $sk_{\mathsf{ABE}}$ of a requesting party.
Any attribute that the requesting party was authorized to access at any time in the past was most likely already retrieved and possibly even persisted locally.
%This is not inherent to our design but a basic fact when it comes to information control.
Consequently, it is not a goal to revoke access to the \emph{current} attributes that were already published.
The primary goal is to prohibit a requesting party from continuously accessing up-to-date attribute information \emph{in the future}.

Our revocation scheme is enforced through attribute versioning.
As elaborated in the previous sections, an attribute record is encrypted using a tag that is a concatenation of the attribute name and version.
When access of a requesting party to an attribute is revoked, we simply increment the attribute version.
Then, we again publish the encrypted attribute value to the name system.

Any other requesting parties also authorized to access the same attribute must be issued new user keys containing updated tags.
The updated keys are published under the same respective labels $rnd$ and can be resolved if needed.
Using this approach we can limit the amount of re-generated user keys to the number of requesting parties that share one or more attribute authorizations with the requesting party that had its access revoked\footnote{As opposed to re-bootstrapping the whole ABE scheme and issuing/publishing new keys for all RPs}.
Another advantage of this approach becomes evident when taking the first authorization of an RP into account:
Initially, it suffices to create a new user key with the current attribute versions and transfer it to the RP.
As the ciphertext does not need to be updated in this case, the attribute records currently in the name system can then instantly be decrypted by the RP.
We define our $\mathbf{Revoke}()$ procedure in Algorithm~\ref{algo_revoke}.

\IncMargin{1em}
\begin{algorithm}
  \SetKwData{Left}{left}\SetKwData{This}{this}\SetKwData{Up}{up}
  \SetKwFunction{Union}{Union}\SetKwFunction{FindCompress}{FindCompress}
  \SetKwInOut{Input}{input}\SetKwInOut{Output}{output}
  \Input{A ticket $t_{\mathsf{rp}}$ issued to RP}
  %\Output{-}
  \BlankLine
  \For{each attribute $a$ of $t_{\mathsf{rp}}.user$}{
    \If{$a.name \in t_{\mathsf{rp}}.names$}{
      a.version++\;
      $\mathbf{Store}(t_{\mathsf{rp}}.ID_{\mathsf{user}}, a)$\;
    }

  } 
  \For{each ticket $t \neq t_{\mathsf{rp}}$ issued by $t_{\mathsf{rp}}.ID_{\mathsf{user}}$}{
    \If{$\emptyset \neq t.names \cap t_{\mathsf{rp}}.names$}{
      $A_{t}$ $\leftarrow$ \{$a \mid a \in A \wedge a.name \in t.names$\}\;
         $\mathbf{Authorize}(t.ID_{\mathsf{user}}, t.ID_{\mathsf{rp}}, A_{t}, msk_{\mathsf{ABE}})$\;
       }
     }
  \caption{Revoke}\label{algo_revoke}
\end{algorithm}\DecMargin{1em}

\subsection{Implementation Considerations}
\label{sec:implications}
As mentioned above, the concrete choice of ABE and name system used in an implementation partially determines the security properties of \thething{}.
In the following we discuss the most important aspects that are relevant and must be considered by implementers.

\subsubsection{Name System}
%TODO Grothoff et al here
While theoretically all name systems are suitable for our design, practically only name systems with strong security guarantees are reasonable choices for building a decentralised IdP.
An insufficiently resilient name system might be subject to denial of service attacks, rendering the IdP useless. 
Also, bulk collection and enumeration of attributes is unwanted, as it exposes organizational and/or trust relationships.

Grothoff et al.~\cite{ns2018} have categorized state of the art name systems according to their security properties including integrity and availability with respect to strong attacker models.
According to the study, Namecoin and GNS exhibit security properties absent in most other name systems, such as resistance to man-in-the-middle manipulation, request and response privacy and censorship resistance.
The authors conclude that the choice of name system is -- in addition to security considerations -- depending on organizational aspects of the ecosystems.

%The integrity of records in a namespace can be ensured by having the namespace owner provide a digital signature along with the resource records.
%Most name systems follow the same approach are as consequently equally suitable for \thething{} in this regard.
%Namecoin additionally uses a consensus-based ledger to ensure the integrity of name registrations.
%
%When it comes to availability, it is important to consider not only the technological but also the organizational architecture of the name system.
For instance, DNSSEC relies on the distributed design of redundant DNS servers as well as caching.
Even more important is the fact that domain names in DNS are highly regulated, making it a semi-centralized system where only the technology is distributed.
This organizational architecture degrades resilience in the face of strong attackers.
Not to mention that DNSSEC further suffers from a design weakness that results in a privacy issue regarding namespace enumeration\footnote{https://dnscurve.org/espionage2.html, accessed 2017/12/26} for which a mitigation was proposed by Goldberg et al~\cite{goldberg2015nsec5}.

In peer-to-peer-based, decentralised name systems such as Namecoin and GNS this is not a problem.
%Instead, such systems struggle with name squatting, where malicious parties squat a large amount of names.
Namecoin is a blockchain-based name system, availability is addressed by having all records replicated by all participants in a local ledger.
No central authorities are required to manage the structure of the name system since integrity is ensured through the consensus mechanism.
Of course, it is trivial to enumerate namespaces in Namecoin due to the nature of a public ledger.
Further, blockchain-based decentralised systems are still quite new and possibly prone to various attacks on the ledger itself~\cite{hijackbtc2017,Giechaskiel2016}.

GNS, on the other hand, is a name system built on top of a distributed hash table (DHT).
It prevents namespace enumeration and also features an efficient response caching mechanism.
GNS is a petname system that inherently mitigates name squatting by not having globally unique names.
Records are generally not considered confidential in a name system as its primary use-case is resource discovery. 
As such, Namecoin and DNSSEC do not protect the contents of resource records or namespaces in any way. 
In GNS, record, query and response data is protected. 
%The blockchain-based design of Namecoin in particular makes this property hard to satisfy, as all information is redundantly stored by all participants while
Grothoff et al.~\cite{ns2018} discuss the respectivce mechanisms to achieve this.
Records are, by default, encrypted using a symmetric key that can be derived from its record label and namespace.
Further, record queries are protected through a ``query privacy'' ultilising a similar approach.
So unlike DNS, for example, queries cannot be trivially observed by third parties.
This prevents an attacker from easily profiling interactions between users and service.
It should be noted that blockchain-based name systems do not suffer from this particular problem as queries are basically just lookups in a local database.

Based on the above, we conclude that peer-to-peer-based approaches, such as Namecoin or GNS, should be preferred over semi-centralized, distributed systems such as DNS.

%While there is a wide variety of name systems today, we take an exemplary look at three name systems: namecoin, a blockchain-based name system, the GNU Name System (GNS) and DNSSEC, the security extensions for the Domain Name System.
%We look at namecoin and GNS because they offer protection against attacks such as client observation on the network and operator level as well as censorship and/or legal attacks~\cite{DPRIV17}.
%While namecoin is, like GNS, also a peer-to-peer-based name system we include it in our discussion because of its relation to NameID and beacuse of the recent surge of blockchain-based system.
%DNSSEC, on the other hand, is the most widely adopted secure name system.

\subsubsection{ABE Scheme}
When it comes to ABE we have to decide between the two major flavors: Key-Policy ABE (KP-ABE) and Ciphertext-Policy ABE (CP-ABE). 
%In most scenarios that deal with access control CP-ABE is suited better.
As discussed by Borgh et al.~\cite{borgh2016attribute,borgh2017employing} the use of CP-ABE is more intuitive than KP-ABE from the point of view of the encrypter.
This is due to the fact that if keys are issued by a third party, it is non trivial for the encrypter to know who has access to the plaintext.
In our case, the encrypter is the same entity as the key issuer so this limitation does not hold.
Furthermore, policies in our proposed system are composed of a single tag\footnote{An attribute name concatenated with a version number}.
Thus, we are not dependent on having the ability to use conjunctive or disjunctive policies in either ciphertext or key material.
Therefore the chosen ABE flavor is independent from the system design.
In fact, the system properties would not significantly change.
The only thing what would differ is the underlying encryption logic that is ideally never exposed by \thething{} anyway.
One could argue that \thething{} would benefit from recent ABE scheme's like FAME \cite{agrawalfame} in terms of performance regarding the cryptographic operations.
But FAME's performance benefits only take effect when policies become more complex.
The reason for that is because in FAME, there is always a fixed number of pairing operations while in other approaches, such as BSW CP-ABE \cite{bethencourt2007ciphertext}, the number of pairings is determined by the complexity of the policy.
In addition to that, the space complexity of BSW CP-ABE ciphertexts and user keys is less than in FAME because it uses two group elements per attribute while BSW CP-ABE uses three group elements per attribute.
FAME uses type-3 pairings, which are more efficient than the type-1 pairings used in the original BSW CP-ABE~\cite{galbraith2008pairings}.
If more complex policies including significantly higher number of attributes are required, the use of the FAME scheme in an implementation should be preferred.
%Boosting performance by getting rid of pairings at all, like in the lightweighted ABE scheme by Yao et al. \cite{yao2015lightweight} is another direction of future work~\ref{sec:future}.

%\clearpage
\section{Implementation}
\label{sec:impl}
We have implemented\footnote{In \nolinkurl{https://gnunet.org/git/gnunet.git} as part of the identity provider subsystem} \thething{} on top of GNS, a name system that is part of the GNUnet peer-to-peer framework\footnote{\nolinkurl{https://gnunet.org}, accessed 11/29/2017}.
Further, we use a slightly modified but functionally equivalent version of the CP-ABE implementation \emph{libbswabe}\footnote{\nolinkurl{http://acsc.cs.utexas.edu/cpabe/}, accessed 2017/29/11}.
%We chose this library because of its availability and, since it is also implemented in C, easy integration into the rest of our C-based implementation.
We chose this implementation because of its general availability and because we presume that variations of this scheme can provide performance improvements.
The use of \emph{libbswabe} can be considered as a performance baseline with regards to our performance evaluations.

In the following section, we discuss what implications our choice of name system and ABE scheme have on the security properties of our implementation.
Additionally, we present and discuss the results of performance tests that we carried out.
Finally, we show how \thething{} can practically be integrated into an OpenID Connect 1.0 Identity Provider (OIDC IdP) to provide a standards-compliant way of using our design. 

\subsection{Security Properties}
Many security properties of our \thething{} implementation stem from GNS:

\textbf{Availability:}
GNS built on top of the $R^5N$~\cite{evans2011r5n} DHT.
$R^5N$ is designed to perform well in restricted-route environments with malicious participants.
\thething{} directly benefits from the strong security guarantees of $R^5N$, such as high resilience and censorship-resistance.
Records are replicated and stored redundantly in $R^5N$ under a key that is generated by hashing the namespace public key with the query name.
Further, GNS is a petname system where users register names in their own local namespace.
This is unlike DNS, for example, since DNS has a global unique root zone managed by a single organization that delegates sub-hierarchies to other organizations.
The petname approach mitigates the name squatting problem where attackers register names in bulk before legitimate users.

\textbf{Integrity:}
Other name systems, such as DNSSEC or Namecoin, sign either whole namespaces or single resource records.
In GNS, record sets are aggregated by label and signed using a key derived from the namespace private key before being published in the DHT.
The DHT has a built-in signature verification ensuring that only valid results are cached and returned.

\textbf{Privacy and Confidentiality:}
Records in GNS are encrypted using a key derived from the query label and namespace public key before they are published into the network.
In a similar fashion GNS realises query privacy.
The namespace private key is derived using the label of a record to sign the records aggregated by label.
The corresponding derived public key is published along with the record, thus allowing any peer to verify the validity of the record and avoid storing corrupted content.
On the other hand, this does not leak information on the namespace owner.
The query key for records under a specific name is constructed by hashing the name and the public key of the respective namespace.
Even if the record data itself is unprotected, which is not the case in our design, a peer in GNS that stores a record or observes a query can only access its contents if the respective name and key are known to that peer.
Finally, we additionally use the CP-ABE library \emph{libbswabe} in accordance with the \thething{} design.

\subsection{Performance Evaluation}
Since our implementation uses GNS, which is built on top of the $R^5N$ DHT, performance is a concern.
We tested our implementation with regards to the following aspects:
\begin{itemize}
  \item Median time to retrieve a user key
  \item Median time to retrieve an attribute
  \item Performance impact of caching in GNS on attribute retrieval
  \item Performance impact of the number of nodes in the network
\end{itemize}

Our test setup consists of a virtual host with 32 vCPUs at 2,3 GHz and 32GB of RAM.
To determine the median time it takes to resolve a key and subsequently an attribute, we bootstrap a GNUnet network $N$.
Before every test run, $N$ is re-bootstrapped to ensure that any caches are purged.
In each run, we repeat one test 10 times.
We define the test to consist of the following steps:
\begin{enumerate}
  \item Randomly choose a node $A$ that acts as a user and a node $B$ that acts as a requesting party from $N$.
  \item $A$ and $B$ create identities in GNS exchange public keys.
  \item $A$ creates and stores a test attribute $a$ and authorizes $B$ to access it.
  \item Simulate an out-of-band handover of the respective authorization ticket $t$ to $B$
  \item $B$ retrieves the ABE user key $sk_{\mathsf{ABE}}$
  \item $B$ retrieves the attribute $a$ and decrypts the attribute value.
\end{enumerate}
Each time the test is repeated, we randomly choose a different node $B \in N$ while $A$ stays fixed.
We measure the time it takes each $B$ to resolve the user key (Step 5) and the attribute (Step 6).
Between each test, we do not tear down or re-bootstrap the network so that we can investigate the impact of caching on retrieval times.
After 10 tests, we tear down the network to conclude the test run.

We execute 1000 test runs to increase the reliability of our dataset.
Further, as we want to investigate the impact of the size of the network, we ran our experiment for $|N|=50$, $|N|=100$, $|N|=150$ and $|N|=200$ nodes, respectively.

We expect the key and attribute retrieval times within a single test run to initially exhibit a high variance.
However, successive attribute resolutions within a single test run by different parties are expected to be faster and show increasingly less variance due to caching kicking in.
We do not expect the same behaviour for the key retrieval.
In regards to the network size, we expect it to have a negative influence on retrieval times and the respective variance with increasing node count.  

\subsubsection{Results}
\begin{figure}[h]
  \centering
  \includegraphics[width=0.45\textwidth]{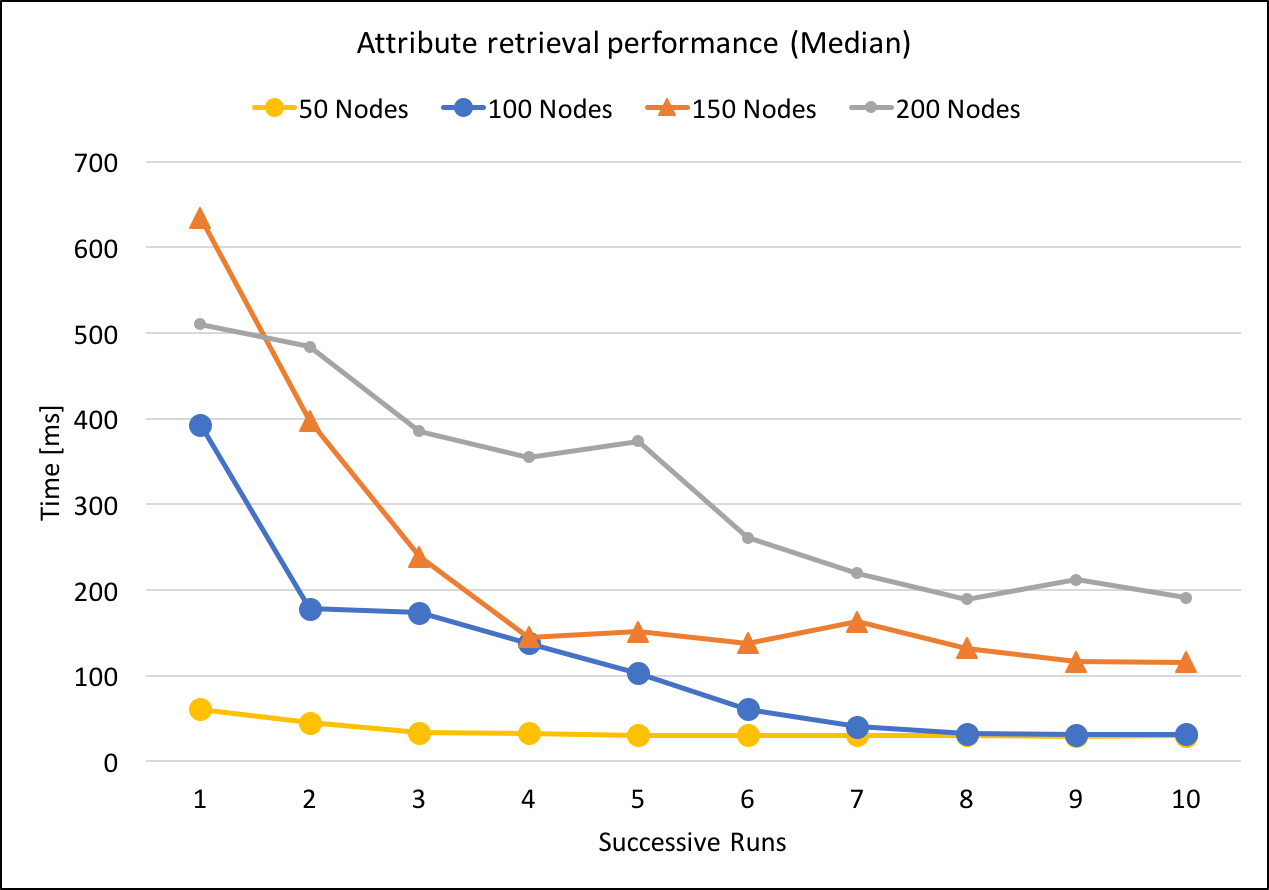}
  \caption{Median attribute retrieval times across all test runs for network sizes of 50, 100, 150, and 200 nodes.}
  \label{fig:nodes_perf}
\end{figure}

Figure~\ref{fig:nodes_perf} is a comparison between median attribute retrieval times across all test runs in differently sized networks.  
The data suggests that the median times for attribute retrieval increases with the size of the network.
At the same time, the retrieval times appear to converge, albeit slower with increasing node count.

\begin{figure}[h]
  \centering
  \includegraphics[width=0.5\textwidth]{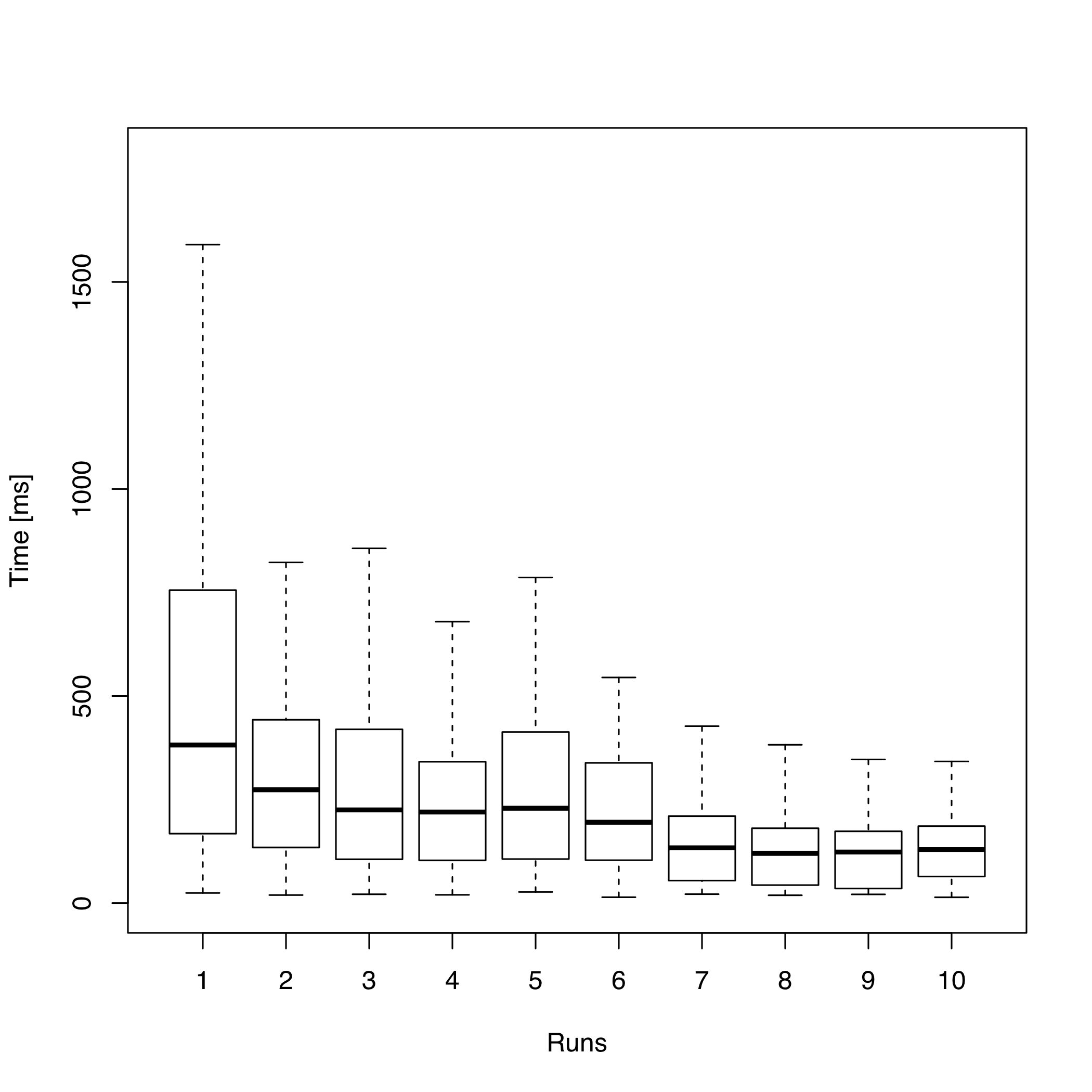}
  \caption{User key retrieval performance of user keys for a network size of 100 nodes.}
  \label{fig:key_perf}
\end{figure}

In Figure~\ref{fig:key_perf}, we can see that the time it takes to resolve a user key varies with a median of around 200 ms.
As expected, the variance is quite high throughout all 10 successive tests across all test runs.
Since we randomly choose our nodes, performance largely depends on the routing and topological distance between $A$ and $B$.
Our experimental setup consists of a clique topology.
However, considering no caching can be leveraged due to the individuality of the query, the observed performance is still practical.
This observation matches with our expectations and suggests that the initial user key after authorization should be transferred to the requesting party out-of-band and not via the name system.
The fact that the initial resolution exhibits a particularly high variance and median retrieval time supports this as well.

\begin{figure}[h]
  \centering
  \includegraphics[width=0.5\textwidth]{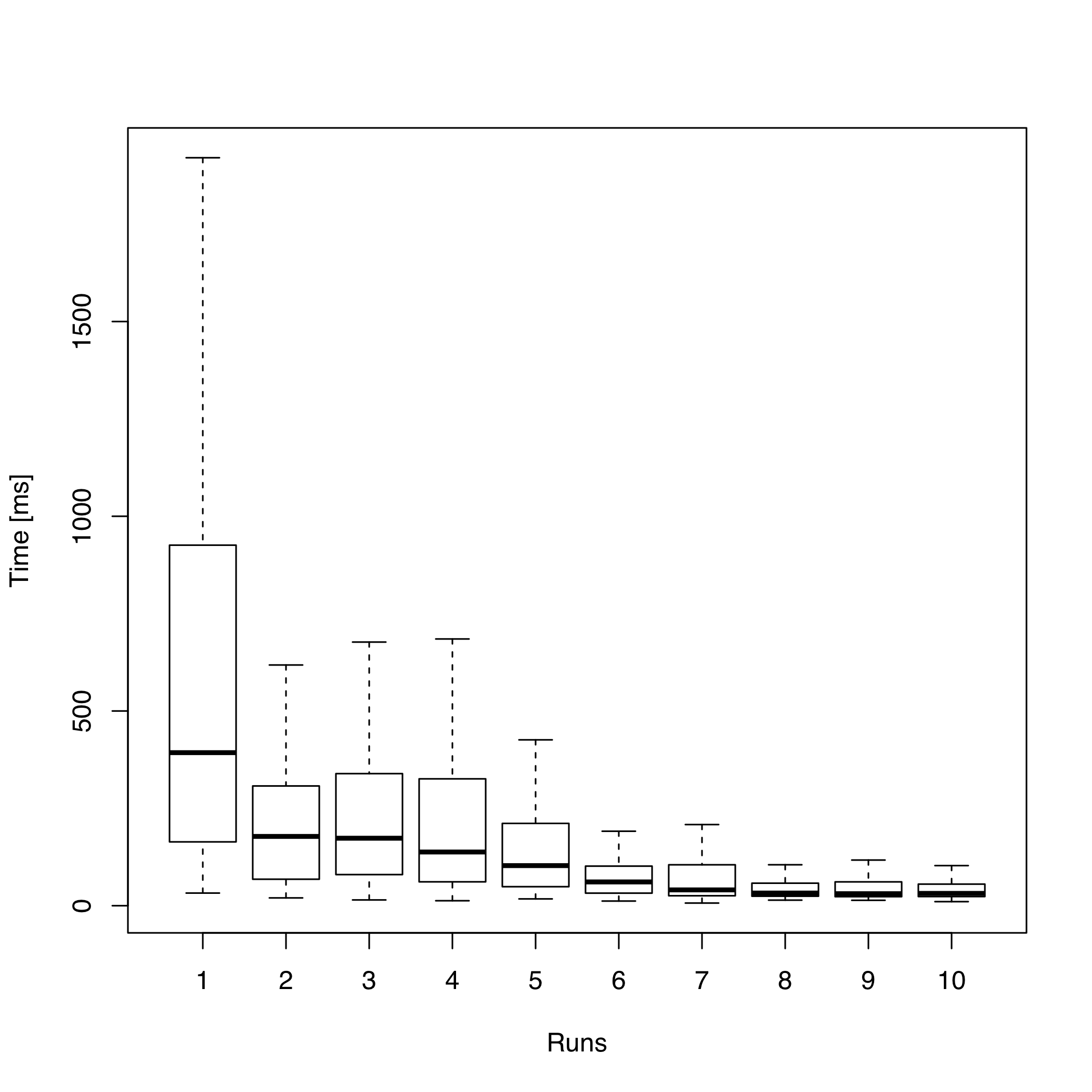}
  \caption{Attribute resolution performance for a network size of 100 nodes.}
  \label{fig:attr_perf}
\end{figure}

In Figure~\ref{fig:attr_perf}, we can see that the retrieval times for attributes also initially exhibit a high variance.
But, retrieval times quickly converge to low median times at less than 100ms with a low variance.
This dataset nicely illustrates the effects of attribute caching in the name system in our implementation.
After the first few requesting parties are authorized to access attributes and have resolved the respective records, resolution times improve greatly.
This confirms our expectation that we can leverage caching of queries and responses in GNS for attributes and it impacts the systems performance positively.

\subsubsection{Discussion}
Our results show that the implementation quickly converges into a reasonably well performing system.
In asynchronous use cases, where user data is retrieved by requesting parties without user interaction after an initial authorization flow, resolution times of up to 100ms are acceptable.
Increasing attribute counts should not negatively impact resolution performance as attributes can be resolved in parallel.

However, since key resolution will not benefit from caching in the current implementation, we recommend an initial out-of-band transfer of the key as part of a authorization protocol such as the OIDC authorization code flow.
Further, keys may change from time to time, in particular due to a revocation initiated by the user. 
When the attributes are required for processing it might already be too late and the attribute can no longer be decrypted with the old key.
It is reasonable for requesting parties to regularly resolve their respective key records, rather than only resolve the updated key.

\subsection{OpenID Connect Integration}
As discussed in our design, an out-of-band exchange of the authorization ticket and possibly even the user key $sk_{\mathsf{ABE}}$ can be done using an authorization protocol.
Instead of proposing our own protocol, we show that our implementation can be abstracted through an HTTP-based OIDC 1.0 compliant authorization flow~\cite{website:oidc}.
Like the specification, in the following we assume that all HTTP exchanges between the user and the service are secured using TLS server authentication.
\begin{figure*}
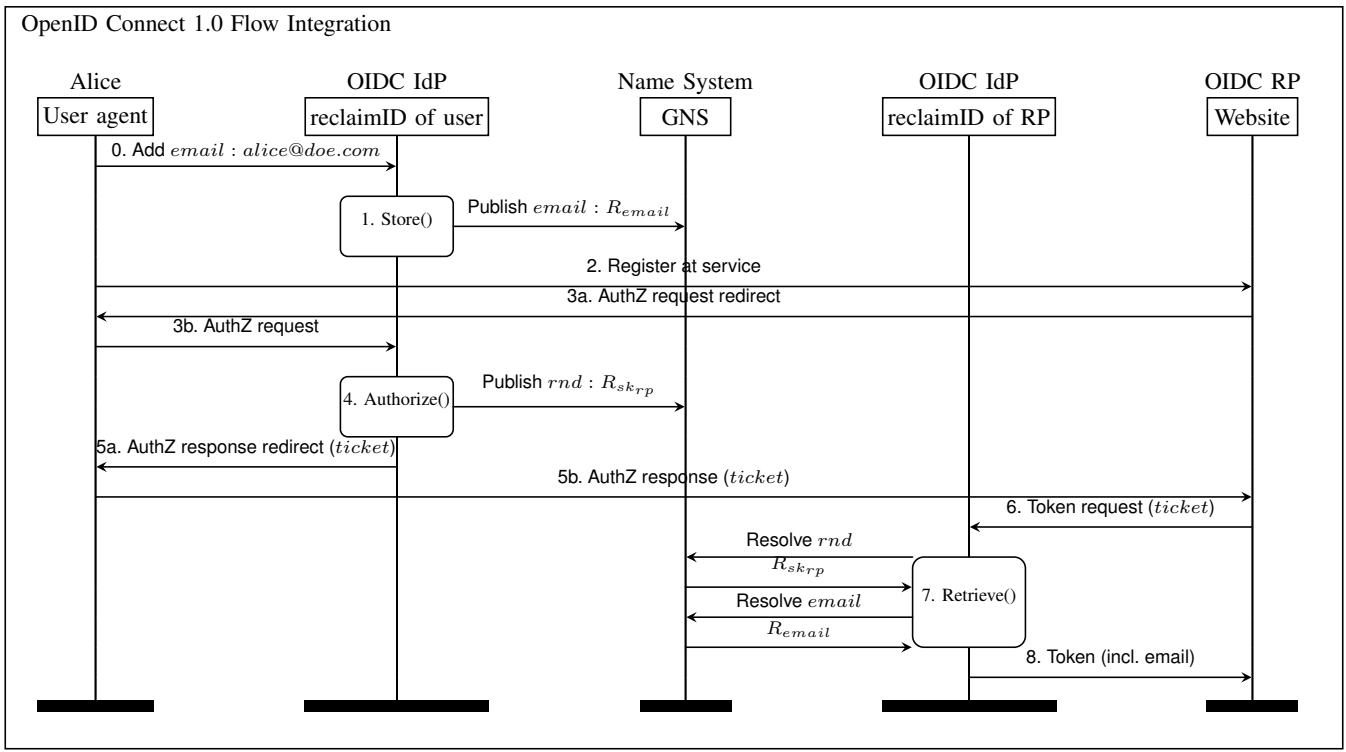

  \centering
  \begin{msc}{OpenID Connect 1.0 Flow Integration}
    \setmscvalues{small}
    \declinst{usr}{Alice}{User agent}
    \setlength{\instdist}{2cm}
    \declinst{gid1}{OIDC IdP}{\thething{} of user}
    \setlength{\instdist}{2cm}
    \declinst{gns1}{Name System}{GNS}
    \setlength{\instdist}{2cm}
    \declinst{gid2}{OIDC IdP}{\thething{} of RP}
    \setlength{\instdist}{2cm}
    \declinst{ws}{OIDC RP}{Website}
    \mess{\textsf{\scriptsize{0. Add $email: alice@doe.com$}}}{usr}{gid1}
    \nextlevel
    \referencestart{store}{\scriptsize{1. Store()}}{gid1}{gid1}
    \nextlevel
    \mess{\textsf{\scriptsize{Publish $email: R_{email}$}}}{storeright}{gns1}
    \nextlevel
    \referenceend{store}
    \nextlevel
    \mess{\textsf{\scriptsize{2. Register at service}}}{usr}{ws}
    \nextlevel
    \mess{\textsf{\scriptsize{3a. AuthZ request redirect}}}{ws}{usr}
    \nextlevel
    \mess{\textsf{\scriptsize{3b. AuthZ request}}}{usr}{gid1}
    \nextlevel
    \referencestart{auth1}{\scriptsize{4. Authorize()}}{gid1}{gid1}
    \nextlevel
    \mess{\textsf{\scriptsize{Publish $rnd: R_{sk_{rp}}$}}}{auth1right}{gns1}
    \nextlevel
    \referenceend{auth1}
    \nextlevel
    \mess{\textsf{\scriptsize{5a. AuthZ response redirect ($ticket$)}}}{gid1}{usr}
    \nextlevel
    \mess{\textsf{\scriptsize{5b. AuthZ response ($ticket$)}}}{usr}{ws}
    \nextlevel
    \mess{\textsf{\scriptsize{6. Token request ($ticket$)}}}{ws}{gid2}
    \nextlevel
    \referencestart{retr1}{\scriptsize{7. Retrieve()}}{gid2}{gid2}
    \mess{\textsf{\scriptsize{Resolve $rnd$}}}{retr1left}{gns1}
    \nextlevel
    \mess{\textsf{\scriptsize{$R_{sk_{rp}}$}}}{gns1}{retr1left}
    \nextlevel
    \mess{\textsf{\scriptsize{Resolve $email$}}}{retr1left}{gns1}
    \nextlevel
    \mess{\textsf{\scriptsize{$R_{email}$}}}{gns1}{retr1left}
    \referenceend{retr1}
    \nextlevel
    \mess{\textsf{\scriptsize{8. Token (incl. email)}}}{gid2}{ws}
  \end{msc}
  \caption{An example authorization flow and attribute retrieval integrated into OpenID Connect (OIDC). Protocol steps 1,2,4,5 and 8 include the standard OIDC authorization Code Flow. Steps 3, 6 and 7 are interactions between the respective local \thething{} and GNS components.}
  \label{fig:oidc_proto}
\end{figure*}
In traditional OIDC deployments, a single service serves well defined endpoints to users and requesting parties.
As our implementation is a decentralised service any participant can take both the role of a user as well as the role of a requesting party.
For this reason, all participants run a local \thething{} instance that exposes the respective OIDC endpoints.

Let us consider our original social network use case illustrated as OIDC flow in Figure~\ref{fig:oidc_proto}:
We assume that a user Alice manages her user attributes -- in particular her email address -- using \thething{} through a web frontend running on her local machine (0).
Consequently, the email record $R_{sk_{rp}}$ containing Alice's address is stored in GNS (1).
She registers to a social networking service at a website (2).
The website offers a ``\thething{}'' button that -- when pressed -- initiates an OIDC authorization code flow in which the service requests access to Alice's ``email'' attribute.
Alice presses the button and her browser is redirected to the OIDC authorization endpoint that is exposed by her local \thething{} installation (3a).
Alice consents to the authorization request (3b) and the authorization procedure as defined in Section~\ref{sec:authorization} is executed (4) that stores the user key $sk_{rp}$ in the record $R_{sk_{rp}}$.
Next, the browser is redirected back to the website (5a,5b) along with an ``authorization code''.
We use the OIDC authorization code to piggyback a reclaimID ticket that includes the $rnd$.

The service exchanges the code at the OIDC token endpoint (6).
This request triggers the retrieval of Alice's email attribute as defined in Section~\ref{sec:retrieval} (7).
The email attribute is wrapped inside a JSON Web Token and returned in the OIDC token response (8).
The response additionally contains an opaque access token that can be used against the OIDC userinfo endpoint.
When Alice is offline, a request to the userinfo endpoint simply reuses the ticket obtained in (5b) to obtain fresh user attributes from \thething{} (6).

\subsubsection{GNS naming and OpenID Connect}
Unlike DNS, GNS is a petname system.
In DNS, there is a global unique root zone managed by a single organisation.
The petname property in GNS comes in handy since the local IdP services for the user and the requesting party can both be addressed using, e.g. ``identity.gnu''.
For each entity, we can assume that the system is configured to map this name to the respective local hosts.
In our design, this host also runs the respective local \thething{} service.
When building HTTP-based authorization protocols on top our design, such as OIDC, this is useful as the specification presumes that there is a single service instance reachable under one domain name. 
Using GNS, this is actually the case while at the same time the service behind the name is decentralised.

%\clearpage
\section{Related Work}
Work related to ours mainly targets decentralised and user-centric identity management. As the following discussion shows, this does not necessarily include a decentralised IdP, which we claim is an essential aspect in a truly user-centric system.

DP5~\cite{borisov2015dp5} is a privacy-friendly presence notification service that through the use of asymmetric pairing functions and pseudo-random functions provides a secure and private personal information retrieval (PIR).
While DP5 addresses the most relevant privacy and security issues, it does not address availability as \thething{} does with a redundant DHT.
A powerful attacker might coerce a DP5 service provider to discontinue its services without having to deal with any of the users trying to share information. 
Participants in \thething{} are not individually protected against such an attack, but the lack of central service instances mitigates the collapse of the whole system.

CONIKS~\cite{melara2015coniks} is an approach originally designed for user-centric key transparency. 
The authors propose a system that does not require a single, centralised third-party to monitor mappings from names to keys, such as from a domain name to its corresponding certificate.
Rather, CONIKS allows users and services to participate in a privacy-preserving protocol that allows them to audit providers of such mappings for non-equivocation.
In contrast to \thething{}, CONIKS relies on centralised IdPs that serve the respective key material. While misuse of those IdPs can be detected, COINKS does not prevent attacks on these central services. This issue is directly addressed by the authors behind ClaimChain~\cite{kulynych2017claimchain}.

ClaimChain also primarily addresses the key verification use-case through the use of hash chains and cross-referencing.
Similar to \thething{}, this approach mitigates the need for trusted centralised IdPs and replaces them with a decentralised protocol and data structure.
While their design proposes flexible, decentralised data structures, ClaimChain does not address the underlying transportation of such.
This is in particular problematic with regards to offline access of claims, i.e. situations where offline users cannot be addressed directly by a requesting party. The authors of ClaimChain state that in this case, claims can be stored at online services for users to interact with them, which would again introduces a centralised component. Our approach addresses this issue by also covering the transportation layer and respective protocols for offline access of attributes.

NameID~\cite{website:nameid} is a recent decentralised IdM approach that uses identities and attributes located in the Namecoin\cite{website:namecoin} blockchain.
Those identities and attributes are in complete control by the respective user and cannot be edited or deleted by the IdP.
As a consequence, all identity attributes in NameID are inherently self-issued, meaning that there is no third party authority issuing or certifying attributes.
Users authenticate by providing proof-of-possession of the respective wallet private key a claimed identity is associated with.
While NameID fully decentralises attribute management, this comes at the cost of privacy.
All identities and attributes are managed in plain text in the Namecoin blockchain and are thus publicly readable to anybody.
We address this issue with \thething{} by a cryptographic access control on attributes using ABE.

Another approach to decentralised IdM by Schanzenbach et al.~\cite{Schanzenbach2016} is also using a decentralised name system as backend for self-issued identity attributes.
Unlike NameID the authors also show how the IdP service itself can be decentralised but their implementation does not feature OpenID Connect compatibility.
Identity attributes that are shared with a requesting party are aggregated and published a single resource record per requesting party. 
However, most name systems are quite unresponsive to changes in namespaces and thus heavily rely on caching which most likely negatively impacts the systems performance.
%References here?

The approach taken by uPort~\cite{website:uPort} is similar to ours in that it is focused on self-issued (in uPort called ``self-sovereign'') identity data stored in a decentralised system.
While we use the name system's distributed storage (specifically in form of a DHT in case of GNUnet), uPort stores public profile data in IPFS~\cite{website:IPFS}. 
Identity data is managed using smart contracts on the Ethereum~\cite{website:ethereum} blockchain.
Currently this approach stores identity data in plain text in the IPFS which raises the same privacy concerns as NameID.
uPort further requires a central service to share private identity attributes between user and requesting party -- something that uPort solves querying the decentralised name system.

\section{Conclusion and Future Work}
\label{sec:future} 
We introduced \thething{}, a decentralised service for self-sovereign identity management.
\thething{} differs from existing approaches in that it combines three main aspects: user-managed attributes without a central party, the complete dissolution of a central IdP into a decentralised query protocol, and privacy-preserving features such as ABE-based access control to sensitive attribute data.

By a practical implementation of \thething{} based on the name system GNS, we have shown that the approach is valid and achieves the functional requirements of an identity management system.
In a series of experiments we evaluated the performance and scalability of our system with the result that in its current state, the implementation of \thething{} is able to serve small to medium applications with up to a few hundreds of participants in production.
Performance optimizations and deployments to large scale testbeds such as PlanetLab are future work.

Through the use of ABE, \thething{} provides an access control layer to user attributes which would otherwise be stored world-readable, as shown by related work.
We consider the ability to authorize access to user attributes not only essential for preserving the user's privacy, but also to enable new use cases which are currently ''solved'' by workarounds that again negatively impact privacy.
For instance, particularly in the context of the Internet of Things, devices often need to be authenticated entities that exist in different security domains. 
Such domain-spanning authentication and authorization scenarios are becoming the rule rather than the exception in an environment full of interconnected devices, but the approaches to this challenge still follow the traditional pattern of adding ''trusted'' third parties (e.g. for device vendors) or complex cross-certification-like constructs.
\thething{} allows a more elegant solution to this challenge by providing an inter-domain identity management infrastructure to establish trust relationships directly between entities, e.g. between a specific device of a vendor and an application by another vendor.

%An improvement allowing conjunctive access policies instead of sole attributes would, without questions, have an impact on the decision which ABE scheme to embed, and is part of the future research.

Besides further performance improvements to address near-realtime requirements, our future work will thus address the extension of \thething{} to support privacy-preserving attribute-based credentials.
%on the basis of succinct non-interactive arguments of knowledge.

\section*{Acknowledegments}
This work was developed in the Fraunhofer Cluster of Excellence $\ll$Cognitive Internet Technologies$\gg$\footnote{\nolinkurl{http://www.cit.fraunhofer.de}}

%\clearpage
\bibliography{references} 
\bibliographystyle{ieeetr}

\end{document}